  \documentclass[12pt,preprint]{aastex}

\begin{document}
\title{New Precision Orbits of Bright Double-Lined Spectroscopic Binaries. I:
    RR~Lyncis, 12~Bootis, and HR~6169}

\author{Jocelyn Tomkin}
\affil{Astronomy Department and McDonald Observatory, University of Texas,
    Austin, TX 78712}
\email{jt@alexis.as.utexas.edu}

\and

\author{Francis C. Fekel\altaffilmark{1}}
\affil{Center of Excellence in Information Systems, 
    Tennessee State University, \\
    3500 John A. Merritt Boulevard,
    Box 9501,
    Nashville, TN 37209}
\email{fekel@evans.tsuniv.edu}

\altaffiltext{1}{Visiting Astronomer, Kitt Peak National Observatory,
National Optical Astronomy Observatory, operated by the Association
of Universities for Research in Astronomy, Inc. under cooperative agreement
with the National Science Foundation.}

\begin{abstract}
Radial velocities from the 2.1 m telescope at McDonald
Observatory supplemented with radial velocities from the coud\'e
feed telescope at KPNO provide new precise orbits for the
double-lined spectroscopic binaries RR~Lyn (A3/A8/A6), 12~Boo
(F8IV), and HR~6169 (A2V).  We derive orbital dimensions
($a_1\sin i$ and $a_2\sin i$) and minimum masses
($m_1\sin ^3i$ and $m_2\sin ^3i$) with accuracies of 0.06 to 0.9\,\%.
The three systems, which have $V$ magnitudes of 5.54, 4.83, and 6.42,
respectively, are all sufficiently bright that they are easily
within the grasp of modern optical interferometers and so afford the
prospect, when our spectroscopic observations are complemented by
interferometric observations, of fully-determined
orbits, precise masses, and distances.  In the case of RR~Lyn,
which is also a detached eclipsing binary with a well-determined
orbital inclination ($i = 87\fdg45 \pm 0\fdg11$;
Khaliullin et al. 2001), we are able to determine
the semimajor axis of the relative orbit, $a = 29.32 \pm 0.04$\,R$_\odot$,
primary and secondary radii of $2.57 \pm 0.02$\,R$_\odot$
and $1.59 \pm 0.03$\,R$_\odot$, respectively;
and primary and secondary masses of $1.927 \pm 0.008$\,M$_\odot$
and $1.507 \pm 0.004$\,M$_\odot$, respectively.
Comparison of our new systemic velocity determination,
$\gamma = -12.03 \pm 0.04$ km~s$^{-1}$, with the earlier
one of \citet{k76}, $\gamma = -11.61 \pm 0.30$ km~s$^{-1}$,
shows no evidence of any change in the systemic velocity
in the 40 years separating the two measurements,
a null result that neither confirms nor contradicts the
presence of the low-mass third component proposed by
\citet{kk02}.  
Our spectroscopic orbit of 12 Boo is more precise that that of  
\citet{betal05} but confirms their results about this system.
Our analysis of HR~6169 has produced a major improvement
in its orbital elements.  The minimum masses of the primary and 
secondary are 2.20 $\pm$ 0.01 and 1.64 $\pm$ 0.02 M$_{\sun}$, respectively.
Although all three systems have eccentric orbits, the six components
of the systems are either pseudosynchronously rotating or very nearly
so.

\end{abstract}


\section{INTRODUCTION}
The current generation of ground-based interferometers, such as the 
Palomar Testbed Interferometer (PTI) \citep{cetal99}, the Naval Prototype 
Optical Interferometer (NPOI)\citep{hetal03}, the Infrared Optical 
Telescope Array (IOTA3)\citep{tetal03}, and the Center for High Angular 
Resolution in Astronomy (CHARA) array \citep{tbetal03}, is advancing 
stellar astronomy in a number of ways.  \citet{q01}, for example, reviewed 
the state of optical and infrared interferometry.  One direction of 
progress is the increasing number of spectroscopic binaries that are 
being resolved as visual binaries.  This allows the determination of 
their three-dimensional orbits and the derivation of accurate masses for 
the component stars and distances to the systems, distances that
in many cases are more accurate than those from the {\it Hipparcos} 
satellite.

In recognition of this development we have started a program to
determine substantially improved spectroscopic orbits for bright, field
spectroscopic binaries.  The program has two benefits: the provision of 
new radial velocities and spectroscopic orbits of a quality that 
matches or exceeds the prospective interferometric observations and, for 
some binaries, the detection of the secondary spectrum and measurement of 
secondary radial velocities for the first time.  We now briefly consider 
these two points in turn.

While some interferometric studies, such as that of 12~Boo
\citep{betal05}, include complementary new radial velocities, the usual 
practise is to take the radial velocities for the binary concerned from 
the literature.  The precision of such velocities often falls short of 
that needed to match the interferometric observations.  For 
example, in their recent determination of the three-dimensional orbit of 
the bright spectroscopic binary $\sigma$~Psc, \citet{kl04} had to complement 
their interferometric measurements with radial velocities observed in 
1944 and 1945 \citep{b47}. Their resulting best-fit solution for the 
three-dimensional orbit has rms velocity residuals of 4.8 and 3.6 
km~s$^{-1}$ for the primary and secondary, respectively.  Orbits with 
large velocity residuals are not exceptional because of the generally 
lower resolution and low signal-to-noise ratio of spectra obtained in the 
first roughly three-quarters of the twentieth century,  For example, of 
the first 100 systems in the {\em Eighth Catalogue of the Orbital Elements 
of Spectroscopic Binaries} \citep{betal89}, 63 have orbits that were 
published in 1980 or earlier and 24 have orbits that were published in 
1950 or earlier, long before the advent of radial velocity
spectrometers and charge-coupled device detectors, which can produce 
spectra with very high signal-to-noise ratios.  Similar proportions 
must apply for all 1469 systems in the Catalogue\footnote{ One would like 
to cite the {\em Ninth Catalogue of Spectroscopic Binary Orbits} 
\citep{petal04} here, but at present it is incomplete with respect to 
orbits published since 1989 and so is unsuitable for the point we are 
making.}.  While these proportions will have improved as a result of the 
substantial number of new spectroscopic binary orbits that have been 
published since 1989 \citep{petal04}, most such orbits are for newly 
discovered binaries. 

With respect to the detection of the secondary spectrum, we note that 
without secondary radial velocities and a determination of the secondary's 
spectroscopic orbit, the linear separation between the binary components 
is unknown and the determination of the three-dimensional orbit is 
incomplete.  Increasing the pool of double-lined spectroscopic binaries 
(SB2s) thus increases the number of spectroscopic binaries available for 
fruitful interferometric observation.  In addition, binary systems 
with components of significantly different masses provide the greatest 
constraints on evolutionary models.  Considering that the majority of 
spectroscopic binaries are single-lined spectroscopic binaries (SB1s), 
there is ample opportunity here.  \citet{petal04}, for example, 
found that two-thirds of the spectroscopic binaries in their 
{\em Ninth Catalogue} are SB1s.  (There is no reason to think the 
Catalogue's incompleteness affects this statistic much.)

Our program uses new, high-resolution, red-wavelength spectra obtained with 
the 2.1 m telescope at McDonald Observatory of the University of Texas and
the coud\'e feed telescope at Kitt Peak National Observatory (KPNO).  
\citet{ft04} provided a preliminary description of our program and an initial
list of observed stars, which has now been expanded to over 40 systems.  
These come primarily from a sample of 130 candidate systems obtained by 
searching the {\em Eighth Catalogue} for SB2 systems that could profit 
from renewed spectroscopic observation and SB1 systems with large enough 
mass functions to suggest that high signal-to-noise ratio spectra might 
transform them into SB2 systems \citep[e.g.,][]{sf92}.  The stars are north 
of $-$40\arcdeg\ in declination and generally brighter than $V = 7.5$ mag.

Others have also seen the need for improved radial velocities for 
spectroscopic binaries.  For example, \citet{k05} has successfully applied 
the iodine absorption-cell method for determining very precise radial 
velocities to the measurement of radial velocities of {\em both} 
components in SB2s.  Hitherto, this technique, which uses an iodine 
absorption cell to impose a reference spectrum on the stellar spectrum 
and is notable for its use in the discovery of extrasolar planets, 
has been restricted to the radial velocities of single stars or stars 
with companions of insignificant relative brightness.  His pioneering 
investigation, which was carried out on the Keck~I telescope with the 
HIRES spectrograph, was limited to five objects including a 
radial-velocity standard and two SB2s.  Among the latter was 64~Psc 
(HD~4676), a well-known, bright spectroscopic binary (F8V, $P = 13.8$ 
days) with a three-dimensional orbit determined by \citet{betal99}, 
using their own interferometric observations made with PTI and radial 
velocities from \citet{dm91}.  Konacki's combined fit of his new 
radial velocities and the \citet{betal99} interferometric data 
leads to better-determined orbital and physical parameters for 64~Psc.  
In particular, the rms velocity residual of 24 m~s$^{-1}$, determined 
from his new fit,  
is a striking improvement compared to the rms residual of 810 m~s$^{-1}$ 
(average $|O - C| = 650$ m~s$^{-1}$) given by \citet{dm91} for 
their radial-velocity solution.  Our new velocities are 
not as precise as the iodine-cell velocities
from the Keck~I telescope --- we will see that for 12~Boo,
which has a similar spectral type to 64~Psc, 
the rms residual for our velocities is 110 m~s$^{-1}$ --- but their 
quality is still a good match with that of current interferometric 
observations.  There is thus ample opportunity for radial velocities 
from small and medium-sized telescopes, measured by traditional methods, 
to make a worthwhile contribution.

Here we report new orbit determinations for three bright spectroscopic 
binaries, RR~Lyn, 12~Boo, and HR~6169.  Analysis of our new velocities 
provides significant improvements in the orbital elements of the  
systems.  Indeed, in the case of HR~6169 some of the revisions are 
substantial.  Table~1 gives basic data for the systems; all three 
are known SB2s, have orbital periods near 10 days, and eccentric orbits.  
We now briefly look at each system individually.

\subsection{RR Lyncis}
RR~Lyn has long been known as both an eclipsing and spectroscopic binary.  
Its orbit has a period of 9.95~days and a modest eccentricity, $e = 0.08$.  
The star's peculiar abundances are indicated in its metallic-lined spectral
classification, which reflects primarily the spectral type of the primary.
\citet{am95} classified it as A3/A8/A6, based on its Ca~II K line, its
hydrogen Balmer lines, and its metallic lines, respectively, while \citet{r49}
called the hydrogen lines A7 and the metallic lines F0.  \citet{ketal01}
compared the two components with evolutionary tracks and also used  
their photoelectric photometry on the $WBVR$ system \citep{ketal85} to place 
the components of RR~Lyn in a $(B - V, V - R)$ diagram.  From those results 
they estimated individual spectral types of A6~IV for the primary and 
F0~V for the secondary.  Thus, the primary has already begun to leave the 
main sequence, while the secondary is still ensconced within it.  
The metallic-line status of the primary is not in doubt \citep{p71,k76,lr95}, 
while that of the secondary remains uncertain, although the abundances 
of \citet{lr95} indicate possible metallicism.

\citet{h15} published the first spectroscopic orbit.  More recent, but 
now quite old orbits were published by \citet{p71} and 
\citet{k76}.  Kondo's orbit has the advantages that it sampled the primary 
and secondary radial velocity curves more fully than Popper's and that 
it has a single systemic velocity, while Popper determined different 
systemic velocities for the primary and secondary.

Photoelectric light curves of RR Lyn have been reported by
\citet{h31}, \citet{mk59}, \citet{b60}, \citet{l66}, \citet{lavetal88}, 
and \citet{ketal01}.  The last of these studies used precise 
photometry in the $W, B, V,$ and $R$ bandpasses \citep{ketal85}, which are 
based on but somewhat different from the Johnson $U, B, V,$ and $R$
photometric bandpasses.  \citet{ketal01} showed that the primary
and secondary eclipse depths in $V$ are $\sim 0.37$ and $\sim 0.25$,
respectively, the fractional luminosities, also in $V$, are
$L_1 = 0.7835 \pm 0.0039$, $L_2 = 1 - L_1$, and the orbital
inclination is $87\fdg45 \pm 0\fdg11$.  Both \citet{l66}
and \citet{b74} needed the addition of third light to solve 
the light curves, while \citet{k76}, who also did a photometric
analysis, found no requirement for third light.  This last result
is supported by \citet{ketal01}, who demonstrated that their observations 
in all filters are fitted by the same geometry and that their model 
matches the observations without any need to invoke third light.  In a 
later paper, however, \citet{kk02} presented new evidence for a third body, 
based {\it not} on an analysis of the light curve, but on the times of 
the eclipse minima.

The presence of the third star, which \citet{kk02} suggested is a very 
low-mass, low-luminosity object that contributes an insignificant amount 
of light to the system, is inferred from quasi-periodic oscillations in 
the times of primary and secondary minima.  They proposed an extemely 
eccentric orbit ($e = 0.96 \pm 0.02$) with a rather long orbital period 
of $39.7 \pm 4.2$ years for it.  The 64-year timespan of the published 
photoelectric minima amounts to only one-and-a-half orbital periods, 
however, so the case for the third body cannot yet be regarded as 
conclusive.

\subsection{12 Bootis}
The SB2 system 12~Boo is the only one in our trio that already has a 
determination of its three-dimensional orbit \citep{betal05}.  The two 
components have an orbital period of 9.60~days and are of very similar 
mass, having a secondary to primary mass ratio of 0.97.  The system's 
combined spectral type is F8~IV \citep{r50} or F8~V \citep{getal01}.  
\citet{betal05} found that the primary is leaving the main sequence, and 
with $T_{eff} = 6130 \pm 100$~K it is now marginally {\em cooler} than 
the secondary ($T_{eff} = 6230 \pm 150$~K).

\citet{h14} was the first to determine its spectroscopic orbit.  Newer 
orbital elements have been published by \citet{al76} and \citet{du99}. 
\citet{betal05} presented the newest set of radial velocities as well as 
interferometric observations.  Details for these more recent spectroscopic
orbits, including the present investigation, are summarised in Table~2.  
Prior to its publication, we were unaware of the new investigation of 
12~Boo by \citet{betal05}. Thus, in spite of the element of (unintentional)
redundancy here, our own study of 12~Boo, based on radial velocities 
acquired from 2002 to 2005, is still of interest because our velocities 
turn out to be significantly more precise, allowing us to determine improved 
spectroscopic orbital elements.

\subsection{HR 6169}
Compared to RR~Lyn and 12~Boo, the previous work on HR 6169 has been quite
modest.   This binary has a combined spectral type of A2~V \citep{cetal69} 
or A1~IV \citep{am95}.  The first and, to date, only spectroscopic orbit 
is that of \citet{y20}, which was based on radial velocities from blue 
photographic spectrograms obtained at the Cassegrain focus of the 1.8 m 
telescope at the Dominion Astrophysical Observatory (DAO).  This SB2 system 
has an orbit with a period of 10.56 days and a moderate eccentricity, 
$e = 0.41$.

\section{OBSERVATIONS AND SPECTRUM REDUCTION}
Our observations were collected from 2002 April through 2005 October
at the McDonald Observatory and KPNO, with the majority of 
spectra of all three program stars obtained at McDonald.  The 
McDonald observations were acquired with the 2.1 m telescope, the 
Sandiford Cassegrain echelle spectrograph \citep{metal93}, and a 
Reticon CCD with 27\,$\mu$m$^2$ pixels arranged in a 1200$\times$400 
pixel format.  The spectrograms cover the wavelength range 5700 to 7000~\AA\ 
and have a resolving power of 60,000.  The collapsed one-dimensional 
stellar spectra have a representative signal-to-noise ratio of $\sim$300.  
Each stellar observation was followed by an observation of a 
thorium--argon comparison spectrum.

The observations at KPNO were obtained with the coud\'e feed telescope,
coud\'e spectrograph, and a TI~CCD detector.  All the spectrograms 
were centered at 6430~\AA, cover a wavelength range of about 80~\AA,
and have a resolution of 0.21~\AA\ or a resolving power of just over 
30,000.  The spectrograms typicaly have signal-to-noise ratios of  
200-250.

The data were processed and wavelength-calibrated in a conventional manner 
with the IRAF\footnote{IRAF is distributed by the National Optical 
Astronomical Observatory, which is operated by the Association for 
Universities for Research in Astronomy, Inc., under contract to the 
National Science Foundation.} package of programs. 

\section{RADIAL VELOCITY MEASUREMENT}
The spectra of the 
three systems are double-lined and, at most orbital phases, the secondary 
lines are well separated from their primary counterparts.  In RR~Lyn there 
are numerous lines of both the primary and secondary with the secondary 
lines being distinctly weaker, while in 12~Boo the primary and secondary 
lines are also numerous and are of roughly similar strength.  In HR~6169 
the primary is represented by only a few lines from ionised species, while 
the secondary has numerous lines, mostly from neutral species, which are 
also sharper than those of the primary.  In this system, therefore, at 
the yellow and red wavelengths covered by our observations the 
secondary lines are much {\em better} for velocity measurement than 
those of the primary.  We now outline the procedures used to measure the 
McDonald and KPNO radial velocities.

For the McDonald velocities we proceeded as follows: the wavelengths of 
well-defined primary and secondary lines were measured by fitting Gaussian 
profiles with the SPLOT routine of IRAF, the wavelength differences 
between the measured wavelengths and rest wavelengths \citep{metal66} of 
the lines provided the topocentric radial velocities, telluric O$_2$ lines 
near 6300~\AA\ ($\alpha$ band) and 6900~\AA\ (B band) were measured in the 
same way so as to determine the offset between the stellar spectrum and 
its associated thorium--argon comparison spectrum, the stellar topocentric 
velocities were then corrected by subtracting from them the telluric line 
offset in velocity form, and, finally, the heliocentric correction led to 
the heliocentric radial velocities.  The McDonald velocities are, thus, 
absolute velocities.

The KPNO radial velocities were determined with the IRAF cross-correlation 
program FXCOR \citep{f93}.  The IAU radial velocity standard stars $\beta$
Vir, HR~5694, and HR~7560, which have late-F spectral classes, were used as 
cross-correlation reference stars.  Their velocities of 4.4, 54.4, and 0.0 
km~s$^{-1}$, respectively, were adopted from \citet{setal90}.  In the KPNO 
spectra, lines in the wavelength region redward of 6445~\AA\ are not 
particularly suitable for measurement because most features are blends, 
and there are a number of modest strength water vapor lines.  Thus, the 
radial velocities were determined from lines in the region 6385~\AA\ to 
6445~\AA.  Unfortunately, this 60~\AA\ portion of the spectrum is so small 
that a mismatch, caused by the varying strength of line blends, between 
the spectral type of the program star and that of the cross-correlation 
standard can significantly affect the measured velocity.  Thus, instead 
of cross-correlating this entire 60~\AA\ wavelength region, only the 
wavelength regions around the strongest and least-blended lines were 
cross-correlated.  For our stars with components of late-A or F spectral 
type the list includes five lines, the Fe~{\sc i} lines at 6394, 6412, 6421, 
and 6431~\AA\ plus the Ca~{\sc i} line at 6439~\AA.  The weakness of those
neutral lines in the spectra of early-A stars means that the only 
measurable lines in their KPNO spectrograms are the two Fe~{\sc ii} lines at 
6417 and 6433~\AA, which can provide good velocities if the lines are
narrow enough.  Each cross-corrrelation profile was fitted with a Gaussian
function.

Because different techniques were used to measure the McDonald and KPNO 
radial velocities, the two velocity systems may have slightly different 
zero points.  We will examine this possibility in \S6.  
Table 3 gives the calendar dates, heliocentric Julian Dates and heliocentric 
radial velocities for the McDonald and KPNO observations.

\section{SPECTRAL AND LUMINOSITY CLASSES}
\citet{sf90} identified several luminosity-sensitive and temperature-sensitive
line ratios in the 6430-6465~\AA\ region.  They employed those critical line 
ratios and the general appearance of the spectrum as spectral-type
criteria.  However, for stars that are hotter than early-G spectral class, the 
line ratios in that wavelength region have little sensitivity to luminosity.
Thus, for the A and F stars of our three systems, we have used the entire 
80~\AA\ spectral region of our KPNO observations to estimate just the spectral 
classes of the individual components.  In the 6430~\AA\ region most of the 
strongest features are 
Fe~I lines with the addition of three Fe~II and three Ca~I lines.
The luminosity class may be determined by computing the absolute visual 
magnitude with the {\it Hipparcos} parallax and comparing that magnitude 
to evolutionary tracks or a table of canonical values for giants and dwarfs.

Spectra of our three binaries were compared with the spectra of a number of
A and F type stars primarily from the lists of \citet{am95} and \citet{f97}.  
The reference-star spectra were obtained at KPNO with the same telescope,
spectrograph, and dectector as our binary-star spectra.  To facilitate a
comparison, the spectra of the reference stars were rotationally broadened
and shifted in radial velocity with a computer program developed by
\citet{hb84} and \citet{b85}. 

This spectrum addition enables us to determine a continuum intensity ratio 
of the binary components at 6430~\AA, a wavelength that is about 0.6 of the
way between the central wavelengths of the Johnson $V$ and $R$ bandpasses.  
If the two stars have very similar spectral types, this intensity ratio 
is also the luminosity ratio and, thus, can be converted directly into a 
magnitude difference.  However, if the lines of the secondary are 
intrinsically stronger than those of the primary, as would be the case, 
for example, if the components were main sequence stars of rather different 
spectral types, then the continuum intensity ratio results in a minimum 
magnitude difference. 

\section{ PROJECTED ROTATIONAL VELOCITIES}
We have determined projected rotational velocities from our red-wavelength
KPNO spectra with the procedure of \citet{f97}.  For each star the full 
width at half-maximum of several metal lines in the 6430~\AA\ region was
measured and the results averaged.  An instrumental broadening of 0.21~\AA\
was removed from the measured broadening by taking the square root of the
difference between the squares of measurements of the stellar and comparison
lines, resulting in the intrinsic broadening.  The calibration polynomial
of \citet{f97} was used to convert this broadening in angstroms into a 
total line broadening in kilometers per second.  For A-type stars this line
broadening corresponds to the {\it v}~sin~{\it i} value.  For F-type stars,
macroturbulent broadening has been taken into account.  Following \citet{f97},
for early-F stars a macroturbulence of 5 km~s$^{-1}$ was adopted and removed,
while for mid-F and late-F stars values of 4 and 3 km~s$^{-1}$, respectively, 
were used.

\section{ORBIT DETERMINATION AND RESULTS}
All three systems have eccentric orbits, so we obtained simultaneous
solutions of the primary and secondary radial velocities by the 
Lehmann--Filhes method to determine the orbits.  A preliminary to the 
orbit calculation was the assignment of suitable weights for the 
various velocities.  Typically the primary and secondary spectra are 
of different quality because the components can differ, sometimes markedly, 
as to strength of lines, line width, and number of available lines, resulting 
in primary and secondary velocities that are of different weight.  Also, the 
McDonald velocities, because of the higher resolution by a factor of two
and the substantially greater wavelength coverage of the McDonald spectra, 
resulting in a larger number of available lines, are somewhat more precise 
than the KPNO velocities and therefore have higher weight.  For each 
system there are, thus, four different sets of velocity, each with its 
own weight.  These weights, which are given in Table~3, were chosen so 
as to approximately equalize the products of the weights and the mean 
square velocity residuals for the different sets of velocity.  Once the 
weights were assigned, the velocities were then solved to determine the 
spectroscopic orbits.   

The velocity residuals (Table 3) for the orbital solutions, given to two 
decimal places to facilitate our examination, allow a straightforward 
check on the consistency of the radial velocities from the two 
observatories.  Any systematic difference between the McDonald and KPNO
velocities will manifest itself as a corresponding difference between 
the average residuals, calculated with regard to sign, for the two types 
of velocity, while if they are consistent then their average residuals, 
apart from statistical fluctuations, will be identical and equal to zero.  
A comparison of the residuals for the McDonald and KPNO radial velocities 
is given in Table~4.

Although the differences are all smaller than or similar to their
estimated errors, they all have the same sign, thus suggesting an offset
of order 0.1 -- 0.2 km~s$^{-1}$ between the McDonald and KPNO velocities.  
We attach the most importance, however, to the primaries of RR~Lyn and 
12~Boo because of the six stars in our three systems these two have 
the most well-determined radial velocities, and in both cases
there is no evidence of an offset.  This simple check, therefore, 
indicates the possible presence of an offset of 0.1 -- 0.2 km~s$^{-1}$ 
between the McDonald and KPNO velocities, but the results are inconclusive.
In the absence of clear evidence of an offset, we decided 
not to adjust either set of velocities.  We now look at the results for 
the individual systems.

\subsection{RR Lyncis}
\begin{figure}
\epsscale{0.90}
\plotone{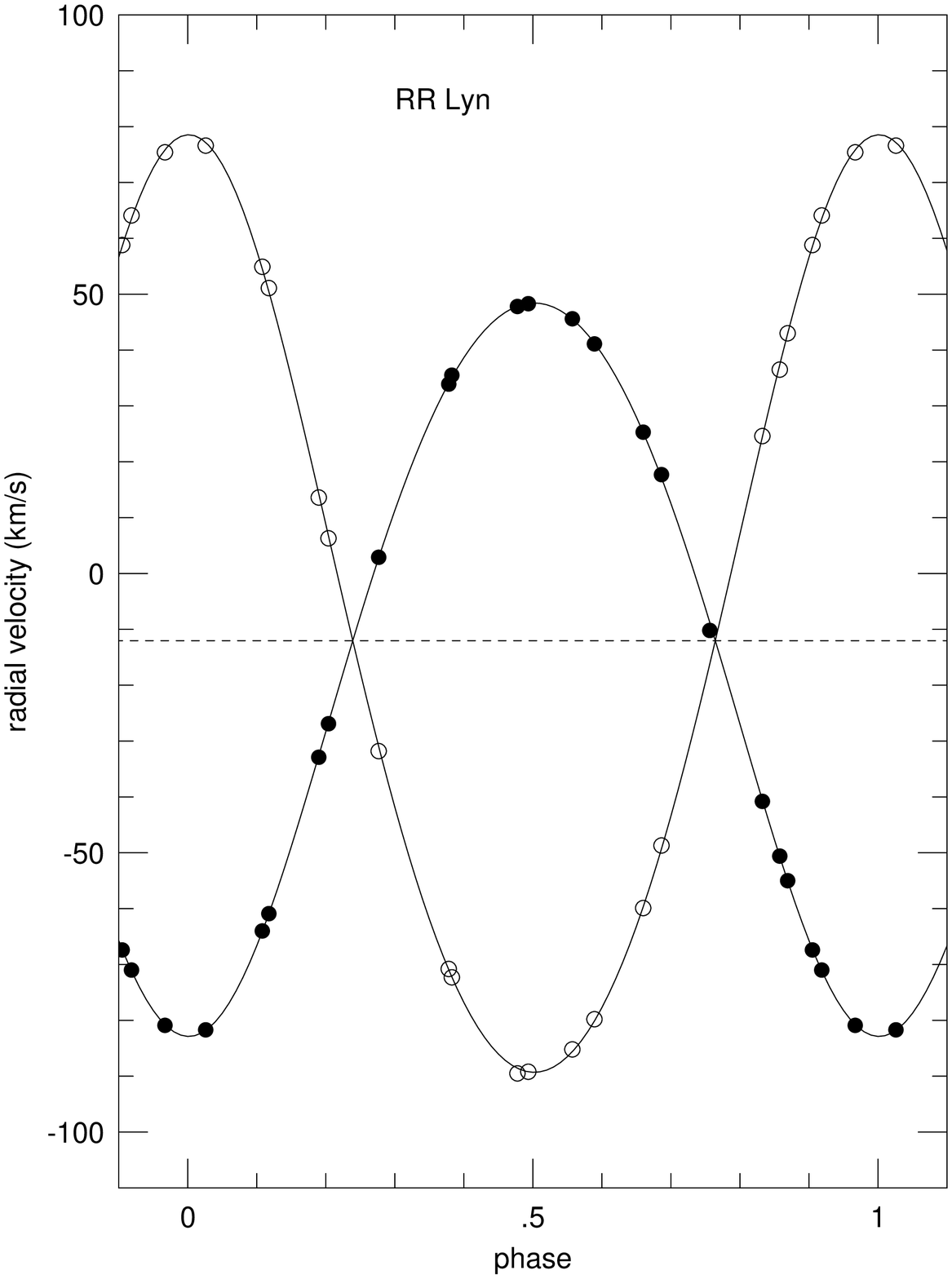}
\caption{
Observed and calculated radial velocity curves for RR~Lyn. Zero phase
is a time of periastron passage.  Filled and open circles are our 
primary and secondary velocities, respectively.
\label{fig1}
}
\end{figure}

Table 5 gives the spectroscopic orbital elements of RR~Lyn
obtained from the solution of our radial velocities.  For comparison,
the elements of \citet{k76} and the photometrically-based elements of 
\citet{ketal01} are also listed.  Figure~\ref{fig1} shows our radial
velocities along with the calculated radial-velocity curves for our 
orbital elements.  
Although the photometrically determined orbital period is more precise 
than our spectroscopically measured value, the difference between the two 
10-day periods corresponds to less than 1 second.  We have retained the 
spectroscopic period for the determination of our spectroscopic orbit, since
adopting the photometric period does not change the orbital elements. 
Examining the solutions in Table~5, one sees that to 
within the uncertainties all three investigations give the same eccentricity 
of 0.08.  This agreement is especially pleasing in the case of 
\citet{ketal01} because their photometric determination of the eccentricity 
is completely independent of our spectroscopic one.  We note that our 
determination of $\omega$ (179\fdg4 $\pm$ 0\fdg6) is the first accurate 
measurement of this element and that to within its uncertainty, it is
$180\arcdeg$, which means the major axis of the true orbit and the
line of nodes coincide.  Also of interest, we find that our and 
Kondo's determinations of $K_1$ and $K_2$ are the same to within the 
uncertainties.  We will compare the two $\gamma$-velocity determinations 
shortly.  The minimum masses are also in agreement, with our uncertainties 
being about 10 times smaller than Kondo's.

As we have seen, RR~Lyn is also a detached eclipsing binary so the
photometric and spectroscopic orbital elements can be combined to provide 
the sizes and masses of the components as well as the linear separation 
between the primary and secondary.  This is done in Table~6.  The new mass 
determinations of $m_1 = 1.927 \pm 0.008$~M$_\odot$ and $m_2 = 1.507 
\pm 0.004$~M$_\odot$ are extremely precise, having uncertainties of 
only 0.4\% and 0.3\%, respectively.  Although our new determinations of 
the masses and radii of the primary and secondary are more precise than 
those of \citet{ketal01}, our values differ only slightly from theirs. 
The new primary and secondary masses are just 0.037\,M$_\odot$ and 
0.017\,M$_\odot$ larger, respectively, than theirs,
while the new primary and secondary radii differ from theirs
by 0.00~R$_\odot$ and 0.01~R$_\odot$, respectively.
We will not rediscuss, therefore, the evolutionary state of RR~Lyn
because our new results would have little impact on the conclusions of 
\citet{ketal01} in this regard.  Instead, we note that insofar as
our new masses and radii confirm theirs, so we support their determinations 
of other derived quantities.  In particular we confirm their photometric 
estimate of the system's distance, 73.5 $\pm$ 2.8~pc, and its age, $1.08 
\pm 0.15$ billion years.  We now look at the question of the
third body in the system.

As already mentioned,  \citet{kk02} proposed the presence of a third
star in an extremely eccentric orbit ($e = 0.96$) with a period of 39.7 years 
as an explanation of variations in the times of minima of the primary and 
secondary eclipses.  The proposed star is of low mass, estimated 
to be 0.10\,M$_\odot$ if its orbit is coplanar with that of the eclipsing 
pair.  Thus, it would make a negligible contribution to the spectrum, but
its presence would cause a regular variation of the systemic velocity of the
eclipsing pair as the center-of-mass of the eclipsing pair orbits the 
center-of-mass of the entire system.  In view of the long interval 
between the observations of \citet{k76} and ours, it is natural to compare 
Kondo's systemic velocity with ours to see if there is a variation.

\begin{figure}
\epsscale{0.80}
\plotone{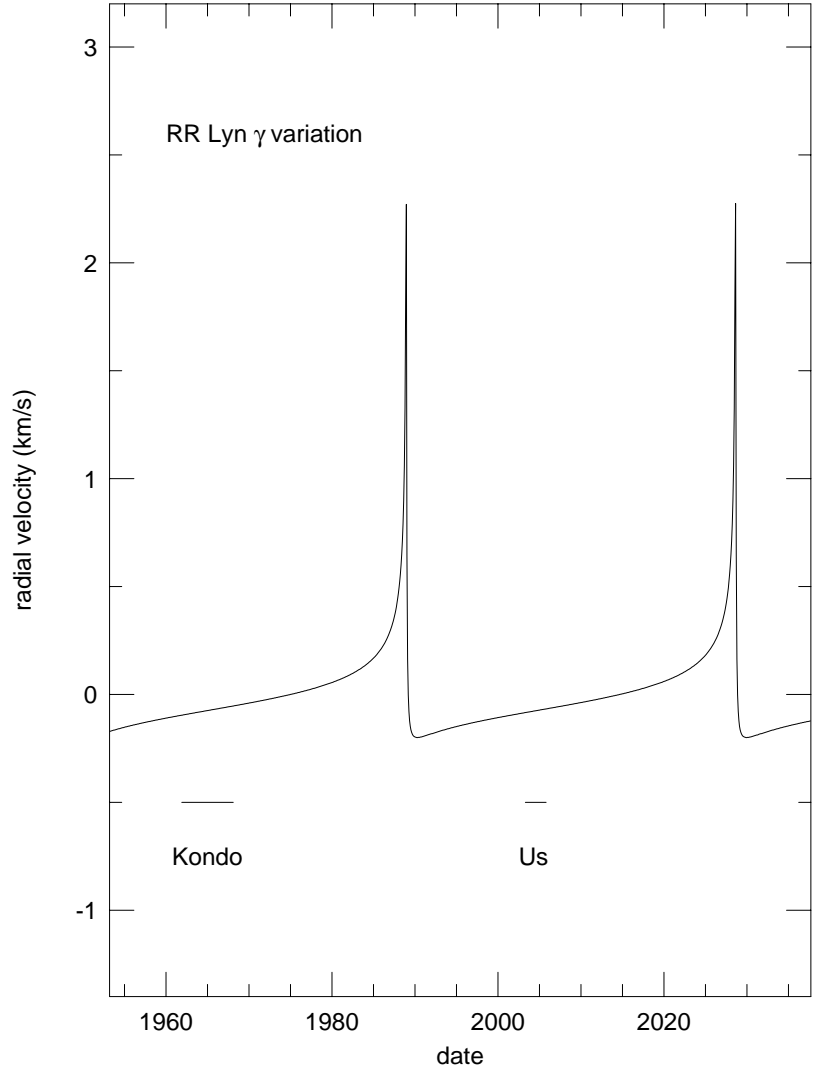}
\caption{
The predicted variation in the $\gamma$-velocity of the
eclipsing pair in RR~Lyn due to the third body proposed
by \citet{kk02}.  The horizontal bars mark the dates spanned
by Kondo's (1976) and our observations.  The tiny predicted
difference, $-0.002$ km~s$^{-1}$, between our and Kondo's
$\gamma$-velocities is well below the threshold of measurement
uncertainty.
\label{fig2}
}
\end{figure}

The difference between our and Kondo's $\gamma$-velocities
is insignificant.  From Table~5 we see that it is $-0.42 \pm 0.30$
with the estimated uncertainty being set by the uncertainty in Kondo's
result.  The observations thus set an upper limit of $\sim0.3$ km~s$^{-1}$ 
on any difference between the two $\gamma$-velocities.  What is the 
predicted difference that would be caused by the proposed third star?  
To answer this question we note that we do {\em not} need to know the 
(unknown) systemic velocity of the entire system; calculation of the 
{\em orbital motion} alone, of the center of mass of the eclipsing pair,
is sufficient to provide the predicted $\gamma$-velocity {\em difference} 
that we want.  \citet{kk02} have provided estimates of all the orbital 
elements needed to calculate the predicted variation of the $\gamma$-velocity
of the eclipsing pair, with the exception of $K_{12}$, and, from the 
information they give, one can infer $K_{12} = 1.31$ km~s$^{-1}$.  We 
can thus calculate the predicted variation of the $\gamma$-velocity of 
the eclipsing pair due to its motion in the proposed 39.7-year orbit, 
and this is shown in Figure~\ref{fig2}.  It is apparent from that figure 
that the predicted difference of $\gamma$-velocities is tiny, only 
$-0.002$ km~s$^{-1}$. 
The smallness of the difference results from the extreme eccentricity of 
the proposed orbit, so the velocity variation outside of periastron
passage is slight, and because, by chance, Kondo's observations
and ours are separated by almost exactly one orbit and thus
are at the same phase.

Our null result, therefore, is consistent with the predicted minute 
change in the $\gamma$-velocity of the eclipsing pair of stars caused 
by the presence of the proposed low-mass third star.  Of course, the 
result is also consistent with the absence of such a body, and so we 
are unable to either confirm or deny its existence.  As can be seen in 
Figure~\ref{fig2}, the extreme eccentricity of the 39.7-year orbit means 
that usually the $\gamma$-velocity of the eclipsing pair changes very 
gradually.  So, during most of the orbit, radial velocities are almost 
impervious to the presence or absence of the third body.  Near periastron 
it is a different story, however, as the third star swoops toward the 
eclipsing pair, and their $\gamma$-velocity suffers an abrupt change 
with a total amplitude of 2.5 km~s$^{-1}$.  At the next periastron 
passage in 2028 one or two precise radial velocities will readily 
confirm or refute the existence of the third star.

\subsection{12 Bootis} 
In their recent investigation of the 12~Boo system \citet{betal05} 
combined their interferometric observations of the visual orbit and 
complementary, new radial velocities to make an accurate determination of 
the system's three-dimensional orbit and the masses and radii of its
components.  They then use these basic parameters to discuss the
system's evolutionary status.  Although our own work duplicates
theirs to some extent, our radial velocities are more precise, so our 
results provide a useful check on the spectroscopic component of their 
investigation.

Table~7 lists the spectroscopic orbital elements from the combined 
interferometric and radial velocity solution of \citet{betal05} with our 
new radial-velocity solution.  The two sets of orbital elements are in 
excellent agreement, and the uncertainties of our elements are usually 
smaller than those of \citet{betal05}.  With the exception of the period, 
for which our observational timespan is quite short, all our elements are 
improved determinations and for four of them ($e$, $K_1$, $K_2$, and 
$\gamma$) the improvement is a factor of two.  There is also good 
agreement between the two $\gamma$ velocities, where we find $\gamma = 
9.588 \pm 0.024$ km~s$^{-1}$, while they find $\gamma = 9.551 \pm 0.051$ 
km~s$^{-1}$, which argues that our and their radial velocities 
have the same zero point.
To make the most of the radial velocities, we also obtained a combined 
solution of our radial velocities and theirs, making no adjustment to 
either set of velocities.  (The weights for our velocities in this
solution are given in Table~3, while the weights for theirs were
0.1 and 0.05 for the primary and secondary, respectively.)

\begin{figure}
\epsscale{1.0}
\plotone{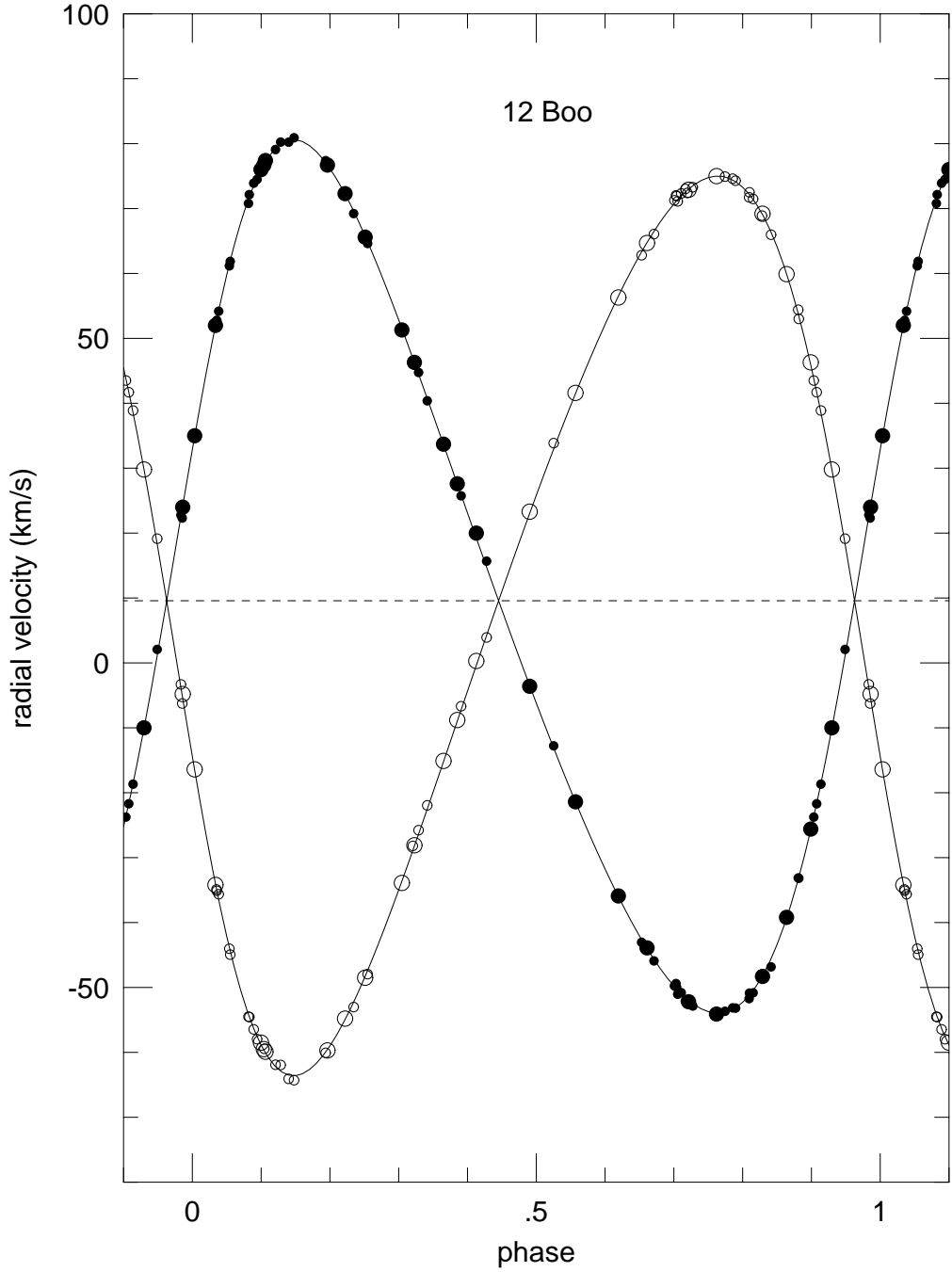}
\caption{
Observed and calculated radial velocity curves for 12 Boo.
Filled and open circles are the primary and secondary velocities,
respectively; large and small circles are the new (McDonald and KPNO)
and CfA velocities, respectively.
\label{fig3}
}
\end{figure}

The elements from this combined spectroscopic solution are also listed 
in Table~7.
They represent an all-round improvement, with the period determination,
in particular, being sharpened thanks to the longer timespan, 1987
to 2005, of the combined observations.
Figure~\ref{fig3} shows our radial velocities and those of \citet{betal05} 
along with the calculated radial velocity curves for the combined solution 
of the two sets.

With the help of the inclination, $i = 107\fdg990 \pm 0\fdg077$,
from the orbit of \citet{betal05} we determine new values of the 
stellar masses and the semi-major axis of the relative orbit.  We find
$m_1 = 1.416 \pm 0.003$\,M$_\odot$,
$m_2 = 1.375 \pm 0.002$\,M$_\odot$,
and $a = 18.611 \pm 0.012$ Gm.
These results certainly confirm those of \citet{betal05}, 
which are
$m_1 = 1.4160 \pm 0.0049$\,M$_\odot$, 
$m_2 = 1.3740 \pm 0.0045$\,M$_\odot$,
but are more precise.

Finally we look at the orbital parallax, $\pi_{orb}$, and distance
of the system.  The orbital parallax is simply $a$ (arcsec)/$a$ (au)
where $a$ denotes both the angular and linear sizes, respectively,
of the true orbit.  The angular size of the orbit, $a = 3.451 \pm 0.018$~mas, 
of course, comes from the visual-orbit determination \citep{betal05}.  
We note, however, that the uncertainty in the angular size of the orbit, 
which is 0.5\% and is much bigger than the uncertainties in the 
determinations of the linear size of the orbit (0.1\% for theirs and 
0.06\% for ours), dominates the uncertainty in the orbital parallax.
Thus, although we have improved the determination of the linear
size of the orbit, there are no corresponding improvements
in the orbital parallax or distance of 12~Boo,
so we content ourselves with restating
the results of \citet{betal05} --- $\pi_{orb} = 27.72 \pm 0.15$\,mas and a
distance to the system of $36.08 \pm 0.19$\,pc.

\subsection{HR 6169}
\begin{figure}
\epsscale{0.90}
\plotone{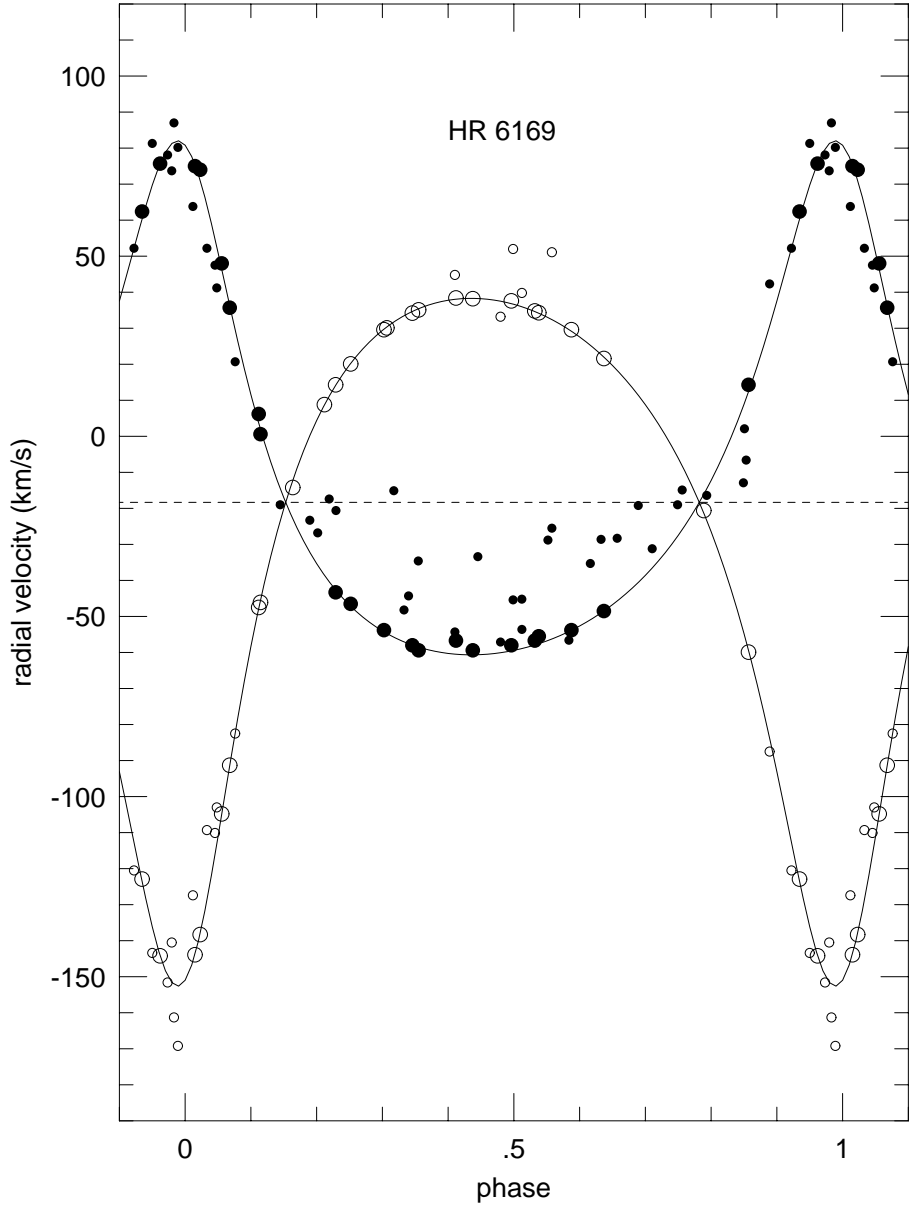}
\caption{
Radial velocities and calculated radial velocity curves for HR~6169.
Filled and open circles are the primary and secondary velocities,
respectively; large and small circles are our and Young's (1920)
velocities, respectively.  The latter set of velocities is shown for
comparison only and was not used in our orbital analysis.  In this system 
the secondary, which has sharper and --- in the red --- more numerous 
lines than the primary, has the more precise velocities.
\label{fig4}
}
\end{figure}

Table~8 lists the orbital elements of HR~6169, derived from the solution
of our radial velocities, as well as the elements obtained by \citet{y20}.  
In Figure~\ref{fig4} our radial velocities and those of \citet{y20} are 
compared with the radial velocity curves calculated from our elements 
alone.  For HR~6169, in the yellow and red region of the spectrum where 
we measured our radial velocities, the only useful
lines representing the primary are the two Si\,{\sc ii} lines at 6347 and 
6371~\AA\ and a smattering of very weak Fe\,{\sc ii} lines, while the 
secondary is represented by lines of neutral and ionised species that are 
both more numerous and sharper.  For this reason the {\em secondary}
velocities are more precise than those of the primary.  It follows that
the secondary velocities have much higher weight than those
of the primary (see Table~3), the secondary radial-velocity curve
is better defined than that of the primary, and $K_2$ is better determined
than $K_1$.   

Comparison of our results with Young's reveals some
very significant discrepancies, especially for the $\gamma$ velocity.
At first sight the large difference of $-8.45$ km~s$^{-1}$ between
our $\gamma$ velocity ($-18.33 \pm 0.09$ km~s$^{-1}$)
and Young's ($-9.88 \pm 0.85$ km~s$^{-1}$) suggests the presence
of an unseen third star, which must have changed the $\gamma$ velocity
of the spectroscopic-binary pair in the 84 years between the two
measurements.  A glance at Figure~\ref{fig4}, however, shows there
is no need to invoke a third star.  Instead, a simpler explanation
of the difference can be found because of a systematic
error in Young's $\gamma$ velocity, caused by blending of the
primary and secondary lines in his spectra.  \citet{y20} remarked that
in his spectra with smaller primary-secondary velocity separations
the primary lines and their secondary counterparts were blended,
so that he was only able to measure a single blend velocity,
which he adopted as the primary velocity.  Inspection of
Figure~\ref{fig4} shows that these blended primary velocities
cluster around the smaller velocity extremum associated with
the nodal passage near apastron.  They mostly fall below the
$\gamma$ velocity and, being dragged upwards and towards it,
are too positive and, not being balanced by the few
velocities that are dragged downwards, their presence in
Young's orbit solution made his $\gamma$ velocity too
positive.  Similar considerations account for his semi-amplitude of 
the primary being too small, $62.41 \pm 1.08$
km~s$^{-1}$ compared with our value of $71.35 \pm 0.38$ km~s$^{-1}$.

Unlike our other two systems, RR~Lyn, which is an eclipsing system,
and 12 Boo, for which interferometric observations have been used
to determine a visual orbit, HR~6169 has no direct determination of
its orbital inclination.  Thus, at present we can only compute
the minimum masses of its two components, which are 2.20 $\pm$ 0.01 and 
1.64 $\pm$ 0.02 M$_{\sun}$ for the primary and secondary, respectively.

\section{SPECTRAL TYPES AND MAGNITUDE DIFFERENCE}
\subsection{RR Lyncis}
\citet{am95} classified the combined spectrum of RR~Lyn as A3/A8/A6,
based on its Ca II K, hydrogen Balmer, and metal lines, respectively,
providing a general starting point for our determination.  
The rapid rotation of most A-type stars plus a variety of abundance 
peculiarities make it difficult to identify a significant number of 
slowly rotating reference stars for that spectral class.  Despite the 
limited choices, identified primarily from the work of \citet{am95}, we 
obtained a reasonable fit to the spectrum of RR~Lyn with HR~3354 
(spectral classes of A3/A5/A7 = calcium/hydrogen/metals \citep{am95}) 
for the primary plus HR~1613 (spectral type of A9~V \citep{am95} and 
[Fe/H] $\simeq$ 0.0 \citep{betal02}) for the secondary.  
While the iron lines of the primary are fitted relatively well by 
HR~3354, the calcium lines of this reference star are too weak.  
Our results for the stellar components, however, are similar to 
those of \citet{ketal01}, who determined spectral types of A6~IV 
plus F0~V from their $WBVR$ photometry and a comparison with 
evolutionary tracks.  The continuum intensity ratio of our 
fit is 0.236, which corresponds to a magnitude difference of 1.6.
This is in rough agreement with the results of \citet{ketal01}, who used
the $WVBR$ photometric system \citep{ketal85} and found a $V$-band luminosity 
ratio of 0.276, which corresponds to a magnitude difference of 1.40.
Our magnitude difference is consistent with the luminosity classes assigned
by \citet{ketal01}. Thus, we adopt their estimated spectral types.

\subsection{12 Bootis}
Guided by previous spectral classifications, we used HR~7560 (F8~V 
\citep{r52,getal01} and mean [Fe/H] = 0.06 \citep{t03}) as a proxy for
both components, but found the lines of the resulting composite 
spectrum fit to be too strong.  However, with HR~5694 (F8~IV-V \citep{jm53} 
and mean [Fe/H] = $-$0.10 \citep{t03}) used for both components we found 
an excellent fit to the spectrum of 12 Boo.  The continuum intensity
ratio of this fit is 0.637.  Because the spectral classes of the two 
stars are essentially identical, this value is also the luminosity ratio 
at 6430~\AA, a wavelength that is about 0.6 of the way between the 
center of the Johnson $V$ and $R$ bandpasses.  The intensity ratio produces 
a magnitude difference of 0.5 with an estimated uncertainty of 0.1.  
Such a difference is in accord with the more precise $\Delta$$V$ 
magnitude of 0.48 determined by \citet{betal05}.  The slightly subsolar
iron abundance is also in agreement with the results of \citet{b90}
and \citet{letal99}, who found [Fe/H] = $-$0.03 $\pm$ 0.09 and 
$-$0.1 $\pm$ 0.1, respectively.

\subsection{HR 6169}
From DAO photographic plates, \citet{p50} determined magnitude differences
for 82 spectroscopic binaries, including HR~6169.  He described the
spectrum of this system as consisting of two narrow-lined early-A stars 
and concluded that both components were dwarfs.  His low-weight magnitude 
difference in the blue region of the spectrum is 0.7 $\pm$ 0.4.

Unlike our results for 12~Boo and RR~Lyn we have had difficulty finding 
a good combination of reference stars to represent the spectrum HR~6169.  
Given the A spectral class of the components, the lines of both stars are 
quite weak in the 6430~\AA\ region with all but one of the lines being 
less than 5\% deep in the combined spectrum.  The primary component has the 
stronger Fe~{\sc ii} lines and weaker Fe~{\sc i} lines and so is earlier in 
spectral class.  From the metal lines our best estimates of those spectral 
classes are A1 for the primary and A7 for the secondary with a continuum 
intensity ratio in the 6430~\AA\ region that corresponds to a minimum
magnitude difference of $\sim$1.  

Alternatively, the spectral types may be estimated from the colors of
the components.  Guided by the above results, we found that the $B-V$ 
colors for stars of spectral types A1~V and A5~V, taken from Table~II of 
\citet{j66}, plus an adopted $V$ magnitude difference of 0.9 
result in the observed $B-V$ = 0.06 from the {\it Hipparcos} satellite 
\citep{petal97}.  We have adopted those spectral types and colors in our 
additional analyses.

\section{CIRCULARIZATION AND SYNCHRONIZATION}
   The two main theories of orbital circularization and rotational
synchronization (e.g.; \citealp{z77,tt92}) disagree significantly
on absolute time scales but do agree that synchronization should occur 
first.  From surveys of B and A stars
\citet{aetal02} concluded that synchronization
occurs faster than circularization, confirming for early-type
stars the theoretical results.
The three systems discussed here have orbital periods between
9.6 and 10.6 days and eccentric rather than circular orbits.  
Since two of the three systems contain A-type stars, the observed 
eccentricities of their orbits are not surprising.  \citet{mm92} 
examined 62 spectroscopic binaries with A-type primaries and periods 
less than 100 days.  They concluded that all systems with orbital 
periods $\lesssim$~3 days have circular or nearly circular orbits.  
They also found that many binaries with periods in the range of 3 to 
10 days have circular orbits, while all those with periods greater 
than 10 days have eccentric orbits.  Thus, the eccentric orbits of 
RR~Lyn and HR~6169 appear to be typical of A-type systems with periods
near 10 days.

\citet{h81} has shown that in an eccentric orbit a star's rotational 
angular velocity will tend to sychronize with that of the orbital 
motion at periastron, a condition called pseudosynchronous rotation.  
With equation (42) of \citet{h81} we calculated pseudosynchronous 
periods for the three systems.

\subsection{RR Lyncis}
For RR~Lyn our {\it v}~sin~{\it i} values are 14.6 $\pm$ 1 and 
11.3 $\pm$ 2 km~s$^{-1}$ for its primary and secondary, respectively.  
The uncertainties are estimated, with that for the secondary being
larger because of the weakness of its lines.  We assume that the orbital
and rotational axes are parallel, so the rotational inclination is  
87\fdg4.
Since it is so close to 90\arcdeg, we adopt our {\it v}~sin~{\it i} values
as the equatorial rotational velocities of the components.  From 
Table~6 the radii are 2.57 and 1.59 R$_{\sun}$ for the primary 
and secondary, respectively.  Those values, combined with the 
pseudosynchronous period of 9.59 days, lead to rotational velocities
of 13.6 and 8.4 km~s$^{-1}$.  Thus, the rotation of the primary is 
consistent with its pseudosynchronous value, while the secondary  
appears to be rotating slightly faster than pseudosynchronous.

\subsection{12 Bootis}
Our projected rotational velocities for 12~Boo, 13.0 $\pm$ 1 and 
10.3 $\pm$ 1 km~s$^{-1}$ for its primary and secondary, respectively, 
are in good agreement with those determined by \citet{du99} and 
\citet{betal00}.  Again we assume that the orbital and rotational axes 
are parallel, so the rotational inclination is 108\arcdeg.  Thus, 
the observed equatorial rotational velocities become 13.7 and 
10.8 km~s$^{-1}$.  The radii of the two components \citep{betal05} and
the pseudosynchronous period of 7.84 days, lead to predicted rotational 
velocities of 15.9 and 12.0 km~s$^{-1}$.  Thus, the primary is rotating 
slightly slower than its predicted velocity, while the secondary may well
be rotating pseudosynchronously.

\subsection{HR 6169}
To determine if the components of HR~6169 are pseudosynchronously 
rotating we must first estimate their radii.
From the {\it Hipparcos} catalog \citep{petal97} the $V$ magnitude and
$B-V$ color of the HR~6169 system are 6.42 and 0.062, respectively.
The {\it Hipparcos} parallax of 6.17 $\pm$
0.087 (mas) \citep{petal97} corresponds to a distance of 162 $\pm$
24 pc.  Although at such a distance there may be a modest amount of
interstellar reddening, the color of the system is consistent with the
spectral type of the primary, and so we have assumed no interstellar
reddening.  As a result, the parallax, the $V$ magnitude, and the
adopted $V$ magnitude difference of 0.9 (\S7.3) were combined to obtain
absolute magnitudes $M_V$ = 0.8 $\pm$ 0.3 mag and $M_V$ = 1.7 $\pm$ 0.3
mag for primary and secondary, respectively.  A $B-V$ color of 0.03
for the primary and a $B-V$ of 0.14 for the secondary \citep{j66} 
were adopted and then used in conjunction with Table~3 of \citet{f96}, 
to obtain the bolometric corrections and effective temperatures of 
the two components.  The resulting luminosities of the primary and 
secondary are $L_{1}$ = 43 $\pm$ 12 $L_{\sun}$ and $L_{2}$ = 17 $\pm$ 
5 $L_{\sun}$, respectively, while the radii are $R_{1}$ = 2.6 
$\pm$~0.4~$R_{\sun}$ and $R_{2}$ = 2.1 $\pm$ 0.3 $R_{\sun}$, respectively.  
The uncertainties in the computed quantities are dominated by the parallax
and magnitude difference uncertainties plus, to a lesser extent, the
effective temperature uncertainty, which is estimated to be $\pm$200~K.

Because the orbit of HR 6169 has a moderately large eccentricity of 
0.414, the pseudosynchronous period of 4.95 days is less than half 
of the orbital period.  With that period and the above radii, we 
obtain pseudosynchronous rotational velocities of 26.6 
and 21.5 km~s$^{-1}$ for the primary and secondary, respectively.
We have determined {\it v}~sin~{\it i} values of 32 $\pm$ 5 and 18 $\pm$ 2
km~s$^{-1}$ for the primary and secondary, where the uncertainties are 
estimated.  The minimum masses of the components are rather large, 
suggesting a relatively high orbital inclination.  A comparison with 
canonical values of absolute magnitudes and masses for dwarfs of the
appropriate  spectral types \citep{g92} suggests an inclination of 
about 70\arcdeg.  Adopting this value, assuming, as has been done for 
the other two systems, that the rotational inclination has the same 
value, produces equatorial velocities of 34 and 19 km~s$^{-1}$.  Thus,
the primary appears to be rotating slightly faster than its 
pseudosynchronous velocity, while the secondary may be rotating 
pseudosynchronously. 
 
\section{SUMMARY}
We have determined new spectroscopic orbits for the bright SB2 systems 
RR~Lyn, 12~Boo, and HR~6169.  The accuracies of the minimum masses 
($m_1\sin ^3i$ and $m_2\sin ^3i$) range from 0.13\% for the secondary 
of 12~Boo to 0.9\% for the secondary of HR~6169, while the accuracies 
of the linear sizes of the relative orbit ($a\sin i$) range from 0.05\% 
for 12~Boo to 0.25\% for HR~6169.  These results show our radial 
velocities provide well-determined spectroscopic orbits. which do 
justice to interferometric measurements of 12~Boo \citep{betal05} and, 
we expect, will also do justice to prospective interferometric 
measurements of RR~Lyn, HR~6169, and other systems in our program.

For RR~Lyn, an eclipsing binary with a well-determined orbital 
inclination, we have determined accurate new masses and radii for its 
components.  The semi-major axis of the relative orbit, $a = 29.32 
\pm 0.04$\,R$_\sun$, plus the distance to the system, $73.5 \pm 2.8$~pc 
\citep{ketal01}, lead to a predicted angular size $a = 1.86$~mas for 
the true orbit and, because $\omega$ ($= 179\fdg4 \pm 0\fdg6$) is 
$180^\circ$ to within its measurement error so the semi-major axis lies 
across the line-of-sight, this will also be the $a$ of the apparent 
orbit.  We find that the expected effect of the unseen third star 
proposed by \citet{kk02} on our and Kondo's (1976) radial velocities 
is so slight as to be well below the threshold of our measurement errors.

We have determined a new spectroscopic orbit for 12~Boo, which confirms, 
and is more precise than, that determined by \citet{betal05} as part of their 
measurement of the three-dimensional orbit of this system.  Our 
investigation of HR~6169 corrects systematic errors in Young's (1920) 
orbit and leads to significant revisions of the minimum masses and 
$a_{1,2}\sin i$.  At the distance of HR~6169, $162 \pm 23$\,pc 
(Perryman et al. 1997), the $a\sin i$ of 22.05 $\pm$ 0.05~Gm corresponds 
to an angular separation of 0.91\,mas.

In conclusion we find that radial velocities observed with
small and medium-sized telescopes and measured by
traditional methods provide a suitable spectroscopic
contribution to the combination of measurements required for
determination of three-dimensional orbits.

\acknowledgments
We thank Eshwar Reddy and Chris Sneden for some helpful tips and
David Doss for his patient guidance, which was essential for the
successful operation of the 2.1-m telescope at McDonald
and its instrumentation.  We also thank Richard Green, former 
director of KPNO, for continuing to make the coud\'e feed telescope 
available.  We appreciate Daryl Willmarth's invaluable assistance with
the KPNO coud\'e feed telescope and our observations.  This research at
Tennessee State University was supported in part by NASA grant NCC5-511
and NSF grant HRD-9706268.

\clearpage



\clearpage

\clearpage

\clearpage

\clearpage

\begin{deluxetable}{lcrccccrc}
\tablewidth{0pt}
\tablecaption{BASIC PROPERTIES OF THE PROGRAM STARS\label{tbl-1}}
\tablehead{
\colhead{} & \colhead{} & \colhead{} & \colhead{Spectral} & \colhead{} 
& \colhead{} & \colhead{Parallax\tablenotemark{a}} & \colhead{Period} 
& \colhead{}    \\
\colhead{Name} & \colhead{HR} & \colhead{HD} & \colhead{Type} & 
\colhead{$V$\tablenotemark{a}} & \colhead{$B-V$\tablenotemark{a}} 
& \colhead{(mas)} & \colhead{(days)} & \colhead{Eccentricity}
}
\startdata
RR Lyn   &   2291  &   44691  & A3/A8/A6 & 5.53  &  0.238  & 12.01 &   9.95  &   0.08    \\
12 Boo   &   5304  &  123999  & F8IV & 4.82  & 0.541 & 27.27 &  9.60  &   0.19    \\
 ...     &   6169  &  149632  & A2V & 6.42  &  0.062  & 06.17 &  10.56  &   0.41    \\
\enddata
\tablenotetext{a}{\citet{petal97}}
\end{deluxetable}


\clearpage

\begin{deluxetable}{lccc}
\tablewidth{0pt}
\tablecaption{
RECENT AND NEW SPECTROSCOPIC INVESTIGATIONS OF 12~BOO\label{tbl-2}}
\tablehead{
\colhead{} &
\colhead{Years of} &
\colhead{Number of} &
\colhead{Mean $|O - C|$}   \\
\colhead{Investigation} & \colhead{Observation} & \colhead{Observations} &
\colhead{(km~s$^{-1}$)}
}
\startdata
Abt \& Levy 1976            &  1966 -- 1971  &  20  &  1.3   \\
De Medeiros \& Udry 1999    &  1987 -- 1992  &  12  &  0.90  \\
Boden et al. 2005           &  1987 -- 2001  &  49  &  0.39  \\
This paper (McDonald and KPNO) &  2002 -- 2005  &  24  &  0.14  \\
\enddata
\end{deluxetable}



\clearpage

\begin{deluxetable}{lrrrrrrrrr}
\tabletypesize{\scriptsize}
\tablewidth{0pt}
\tablecaption{RADIAL VELOCITIES OF THE PROGRAM STARS\label{tbl-3}}
\tablehead{
       &
       &
\colhead{HJD}             &
       &
\multicolumn{2}{c}{Velocity}  &
\multicolumn{2}{c}{Weight}    &
\multicolumn{2}{c}{$O - C$}   \\
 & &
\colhead{$-$2,400,000}        &
 &
\colhead{Pri}             &
\colhead{Sec}             &
\colhead{Pri}             &
\colhead{Sec}             &
\colhead{Pri}             &
\colhead{Sec}            \\
\colhead{Date (UT)} & 
\colhead{Tel} & 
& 
\colhead{Phase} &
\colhead{(km~s$^{-1}$)}    &
\colhead{(km~s$^{-1}$)}    &
\colhead{(km~s$^{-1}$)}    &
\colhead{(km~s$^{-1}$)}    &
\colhead{(km~s$^{-1}$)}    &
\colhead{(km~s$^{-1}$)}
}
\startdata
\multicolumn{10}{c}{RR Lyn} \\
\hline
 2003 Apr 22 & McD &  52,751.635 &  0.3781 &    33.9  &  $-70.8$  &  1.000 & 0.100   & $-0.17$ &   0.16   \\
 2004 Feb 07 & McD &  53,042.843 &  0.6598 &    25.3  &  $-59.9$  &  1.000 & 0.100   &   0.03  & $-0.19$  \\
 2004 Feb 08 & McD &  53,043.803 &  0.7563 &  $-10.2$ &  \nodata  &  0.000 & \nodata & $-1.44$ & \nodata  \\
 2004 Feb 09 & McD &  53,044.812 &  0.8577 &  $-50.6$ &    36.5   &  1.000 & 0.100   &   0.11  & $-0.92$  \\
 2004 Mar 10 & McD &  53,074.760 &  0.8691 &  $-55.0$ &    43.0   &  1.000 & 0.100   &   0.03  &   0.06   \\
 2004 Mar 11 & McD &  53,075.734 &  0.9670 &  $-80.9$ &    75.4   &  1.000 & 0.100   & $-0.12$ & $-0.46$  \\
 2004 Apr 13 & McD &  53,108.648 &  0.2766 &     2.9  &  $-31.8$  &  1.000 & 0.100   &   0.12  & $-0.84$  \\
 2004 Apr 15 & McD &  53,110.647 &  0.4776 &    47.8  &  $-89.5$  &  1.000 & 0.100   & $-0.06$ & $-0.92$  \\
 2004 Apr 24 &  KP &  53,119.647 &  0.3826 &    35.5  &  $-72.3$  &  0.500 & 0.050   &   0.44  & $-0.07$  \\
 2004 Sep 25 &  KP &  53,274.019 &  0.9050 &  $-67.4$ &    58.8   &  0.500 & 0.050   & $-0.13$ &   0.21   \\
 2004 Sep 28 &  KP &  53,276.991 &  0.2039 &  $-26.9$ &     6.3   &  0.500 & 0.050   & $-0.20$ & $-0.43$  \\
 2004 Nov 25 & McD &  53,334.893 &  0.0260 &  $-81.7$ &    76.6   &  1.000 & 0.100   &   0.10  & $-0.57$  \\
 2004 Dec 03 & McD &  53,342.913 &  0.8325 &  $-40.8$ &    24.6   &  1.000 & 0.100   & $-0.24$ &   0.15   \\
 2005 Jan 29 & McD &  53,399.846 &  0.5572 &    45.6  &  $-85.2$  &  1.000 & 0.100   &   0.07  &   0.40   \\
 2005 Apr 28 & McD &  53,488.717 &  0.4934 &    48.3  &  $-89.2$  &  1.000 & 0.100   & $-0.05$ &   0.01   \\
 2005 Apr 29 & McD &  53,489.670 &  0.5892 &    41.1  &  $-79.8$  &  1.000 & 0.100   & $-0.13$ &   0.31   \\
 2005 Apr 30 &  KP &  53,490.636 &  0.6863 &    17.7  &  $-48.7$  &  0.500 & 0.050   &   0.58  &   0.59   \\
 2005 Sep 21 &  KP &  53,634.874 &  0.1898 &  $-32.9$ &    13.6   &  0.500 & 0.050   & $-0.28$ & $-0.69$  \\
 2005 Oct 18 & McD &  53,662.011 &  0.9185 &  $-71.0$ &    64.1   &  1.000 & 0.100   &   0.13  &   0.58   \\
 2005 Oct 20 & McD &  53,663.896 &  0.1080 &  $-64.0$ &    54.9   &  1.000 & 0.100   &   0.10  &   0.38   \\
 2005 Oct 20 & McD &  53,663.990 &  0.1175 &  $-60.9$ &    51.1   &  1.000 & 0.100   &   0.04  &   0.61   \\
\hline
\multicolumn{10}{c}{12 Boo} \\
\hline
 2002 Apr 28 & McD &   52,392.711 &  0.1964 &    76.7  & $-59.7$ & 1.000 & 0.500 &  $-0.01$ & $-0.13$  \\
 2002 Apr 29 & McD &   52,393.750 &  0.3046 &    51.3  & $-33.9$ & 1.000 & 0.500 &  $-0.24$ & $-0.26$  \\
 2003 Mar 21 & McD &   52,719.792 &  0.2512 &    65.6  & $-48.5$ & 1.000 & 0.500 &  $-0.16$ & $-0.21$  \\
 2003 Apr 20 & McD &   52,749.700 &  0.3651 &    33.7  & $-15.1$ & 1.000 & 0.500 &    0.06  &   0.10   \\
 2003 Apr 20 & McD &   52,749.892 &  0.3851 &    27.6  &  $-8.8$ & 1.000 & 0.500 &    0.00  &   0.18   \\
 2003 Apr 21 & McD &   52,750.903 &  0.4904 &   $-3.6$ &   23.3  & 1.000 & 0.500 &  $-0.11$ &   0.27   \\
 2003 Sep 06 & McD &   52,888.618 &  0.8289 &  $-48.3$ &   69.2  & 1.000 & 0.500 &  $-0.11$ &   0.12   \\
 2003 Sep 07 & McD &   52,889.586 &  0.9297 &  $-10.0$ &   29.8  & 1.000 & 0.500 &    0.10  & $-0.04$  \\
 2003 Sep 08 & McD &   52,890.584 &  0.0336 &    52.0  & $-34.2$ & 1.000 & 0.500 &    0.09  & $-0.17$  \\
 2004 Feb 08 & McD &   53,043.967 &  0.0034 &    35.0  & $-16.4$ & 1.000 & 0.500 &    0.07  &   0.14   \\
 2004 Feb 09 & McD &   53,044.932 &  0.1039 &    76.7  & $-59.5$ & 1.000 & 0.500 &  $-0.10$ &   0.16   \\
 2004 Apr 13 & McD &   53,108.880 &  0.7620 &  $-54.1$ &   75.0  & 1.000 & 0.500 &  $-0.16$ &   0.00   \\
 2004 Apr 14 & McD &   53,109.858 &  0.8638 &  $-39.2$ &   59.9  & 1.000 & 0.500 &    0.15  & $-0.07$  \\
 2004 Apr 26 &  KP &   53,121.792 &  0.1063 &    77.4  & $-59.9$ & 0.500 & 0.250 &    0.16  &   0.21   \\
 2004 Jun 12 &  KP &   53,168.659 &  0.9860 &    24.0  &  $-4.8$ & 0.500 & 0.250 &  $-0.20$ &   0.69   \\
 2004 Jun 13 &  KP &   53,169.752 &  0.0998 &    76.0  & $-58.5$ & 0.500 & 0.250 &  $-0.01$ &   0.34   \\
 2004 Sep 12 & McD &   53,260.584 &  0.5570 &  $-21.4$ &   41.6  & 1.000 & 0.500 &    0.01  &   0.11   \\
 2004 Sep 13 & McD &   53,261.583 &  0.6610 &  $-43.9$ &   64.7  & 1.000 & 0.500 &    0.11  & $-0.07$  \\
 2005 Apr 27 & McD &   53,487.876 &  0.2220 &    72.3  & $-54.8$ & 1.000 & 0.500 &    0.07  &   0.16   \\
 2005 Apr 28 & McD &   53,488.844 &  0.3228 &    46.3  & $-28.1$ & 1.000 & 0.500 &    0.03  &   0.11   \\
 2005 Apr 29 & McD &   53,489.708 &  0.4128 &    20.0  &    0.3  & 0.000 & 0.000 &    0.74  &   0.69   \\
 2005 May 01 &  KP &   53,491.692 &  0.6193 &  $-35.9$ &   56.3  & 0.500 & 0.250 &    0.09  & $-0.21$  \\
 2005 May 02 &  KP &   53,492.672 &  0.7214 &  $-52.2$ &   72.9  & 0.500 & 0.250 &  $-0.20$ & $-0.10$  \\
 2005 Jun 11 &  KP &   53,532.796 &  0.8990 &  $-25.6$ &   46.3  & 0.500 & 0.250 &    0.13  &   0.36   \\
\hline
\multicolumn{10}{c}{HR 6169} \\
\hline
 2002 Apr 28 & McD &   52,392.790  &  0.2518  &  $-46.5$  &     20.1  &    0.100 & 1.000 &     0.67  & $-0.14$ \\
 2002 Apr 29 & McD &   52,393.881  &  0.3551  &  $-59.4$  &     35.1  &    0.100 & 1.000 &   $-0.88$ & $-0.33$ \\
 2003 Mar 21 & McD &   52,719.891  &  0.2289  &  $-43.3$  &     14.3  &    0.100 & 1.000 &   $-0.67$ &   0.13  \\
 2003 Apr 20 & McD &   52,749.868  &  0.0678  &    35.7   &   $-91.3$ &    0.100 & 1.000 &   $-0.58$ &   0.10  \\
 2003 Apr 21 & McD &   52,750.881  &  0.1637  & \nodata   &   $-14.2$ &  \nodata & 0.000 &  \nodata  & $-1.99$ \\
 2003 Sep 06 & McD &   52,888.661  &  0.2118  & \nodata   &      8.8  &  \nodata & 1.000 &  \nodata  &   0.11  \\
 2003 Sep 07 & McD &   52,889.664  &  0.3068  & \nodata   &     30.1  &  \nodata & 1.000 &  \nodata  & $-0.26$ \\
 2004 Feb 09 & McD &   53,044.975  &  0.0150  &    75.0   &  $-143.9$ &    0.100 & 1.000 &   $-0.38$ & $-0.18$ \\
 2004 Mar 10 & McD &   53,074.983  &  0.8569  &    14.3   &   $-59.9$ &    0.100 & 1.000 &     1.96  & $-0.53$ \\
 2004 Apr 10 & McD &   53,105.944  &  0.7889  & \nodata   &   $-20.6$ &  \nodata & 0.000 &  \nodata  &   0.76  \\
 2004 Apr 24 &  KP &   53,119.942  &  0.1146  &     0.6   &   $-46.1$ &    0.025 & 0.500 &   $-1.38$ & $-0.58$ \\
 2004 Apr 26 &  KP &   53,121.925  &  0.3024  &  $-53.8$  &     29.6  &    0.025 & 0.500 &     0.47  & $-0.14$ \\
 2004 Jun 14 &  KP &   53,170.838  &  0.9345  &    62.4   &  $-122.9$ &    0.025 & 0.500 &     2.09  &   0.66  \\
 2004 Jun 15 &  KP &   53,171.771  &  0.0229  &    74.0   &  $-138.3$ &    0.025 & 0.500 &     2.95  & $-0.37$ \\
 2004 Sep 11 & McD &   53,259.654  &  0.3456  &  $-58.0$  &     34.2  &    0.100 & 1.000 &   $-0.06$ & $-0.46$ \\
 2004 Sep 12 & McD &   53,260.625  &  0.4375  &  $-59.4$  &     38.2  &    0.100 & 1.000 &     1.26  & $-0.09$ \\
 2004 Sep 13 & McD &   53,261.620  &  0.5318  &  $-56.7$  &     34.8  &    0.100 & 1.000 &     1.23  &   0.16  \\
 2004 Sep 13 & McD &   53,261.685  &  0.5379  &  $-55.5$  &     34.3  &    0.100 & 1.000 &     2.07  &   0.14  \\
 2005 Jan 29 & McD &   53,400.003  &  0.6369  &  $-48.5$  &     21.6  &    0.100 & 1.000 &   $-0.03$ & $-0.38$ \\
 2005 Apr 27 & McD &   53,487.910  &  0.9619  &    75.7   &  $-144.2$ &    0.100 & 1.000 &     0.00  & $-0.06$ \\
 2005 Apr 28 & McD &   53,488.899  &  0.0556  &    48.0   &  $-104.8$ &    0.100 & 1.000 &     1.56  &   0.20  \\
 2005 Jun 10 &  KP &   53,531.732  &  0.1119  &     6.2   &   $-47.5$ &    0.025 & 0.500 &     2.55  &   0.25  \\
 2005 Jun 13 &  KP &   53,534.900  &  0.4119  &  $-56.7$  &     38.4  &    0.025 & 0.500 &     3.79  &   0.34  \\
 2005 Jun 14 &  KP &   53,535.789  &  0.4961  &  $-58.0$  &     37.6  &    0.025 & 0.500 &     1.56  &   0.77  \\
 2005 Jun 15 &  KP &   53,536.754  &  0.5875  &  $-53.8$  &     29.6  &    0.025 & 0.500 &     0.04  &   0.43  \\
\enddata
\end{deluxetable}

\clearpage

\begin{deluxetable}{rlrrc}
\tablewidth{0pt}
\tablecaption{AVERAGE VELOCITY RESIDUALS COMPARISON\label{tbl-4}}
\tablehead{
\colhead{}  & \colhead{} & \multicolumn{2}{c}{Average Residual (km\,s$^{-1}$)} 
& \colhead{Difference} \\
\colhead{System}  & \colhead{Component} & \colhead{McDonald} & 
\colhead{Kitt Peak}   & (km~s$^{-1}$)} 
\startdata
RR Lyn & primary   & $-0.003 \pm 0.031$ (15)   & $ 0.082 \pm 0.178$ (5) 
& $ -0.085 \pm 0.181 $  \\
       & secondary   & $-0.083 \pm 0.139$ (15)   & $ -0.078 \pm 0.227$ (5) 
& $(-0.005 \pm 0.266)$  \\
12 Boo & primary   & $-0.012 \pm 0.028$ (17)   & $-0.005 \pm 0.066$ (6) 
& $ -0.007 \pm 0.072 $  \\
       & secondary   & $ 0.024 \pm 0.038$ (17)   & $ 0.215 \pm 0.134$ (6) 
& $ -0.191 \pm 0.139 $  \\
HR 6169 & primary  & $ 0.473 \pm 0.288$ (13)   & $ 1.509 \pm 0.602$ (8) 
& $(-1.036 \pm 0.667)$  \\
       & secondary  & $-0.106 \pm 0.063$ (15)   & $ 0.170 \pm 0.172$ (8) 
& $ -0.276 \pm 0.183 $  \\
\enddata
\tablecomments{The quoted errors are the standard deviations of the 
averages with the number of residuals in parentheses. The differences 
for the secondary of RR~Lyn and the primary of HR~6169 are in parentheses 
because of the relatively low quality of their velocities.}
\end{deluxetable}
\clearpage

\begin{deluxetable}{lrrr}
\tablewidth{0pt}
\tablecaption{ORBITAL ELEMENTS OF RR~LYN\label{tbl-5}}
\tablehead{
\colhead{} & \colhead{} & \colhead{KKK 2001} & \colhead{}  \\
\colhead{Parameter} & \colhead{Kondo 1976} & \colhead{(light curve)} & 
\colhead{This study}
}
\startdata
$P$ (days)                &  9.945080   &      9.9450740    &      9.945080   \\
                          &  fixed      & $\pm 0.0000007$   & $\pm 0.000059$ \\

$T$ (HJD)\tablenotemark{a} & 2,438,048.97  &  2,444,988.49594 & 2,453,334.634 \\
                          & $\pm 0.02$  &     $\pm 0.00030$ &   $\pm 0.017$   \\

$e$                       &      0.081  &          0.0782   &       0.0793   \\
                          & $\pm 0.006$ &     $\pm 0.0009$  & $\pm 0.0009$   \\

$\omega$ (deg)       &    176.1    &        185        &        179.4   \\
                          & $\pm 5.6$   &     $\pm 5$       &  $\pm 0.6$   \\

$K_1$ (km~s$^{-1}$)       &     65.87   &       \nodata     &       65.65   \\
                          & $\pm 0.46$  &                   &  $\pm 0.06$   \\

$K_2$ (km~s$^{-1}$)       &     83.1    &       \nodata     &       83.92   \\
                          & $\pm 1.3$   &                   &  $\pm 0.17$   \\

$\gamma$ (km~s$^{-1}$)    &   $-11.61$  &       \nodata     &    $-12.03$   \\
                          & $\pm 0.30$  &                   &  $\pm 0.04$   \\

$m_1\sin ^3i$ (M$_\odot$) &      1.88   &       \nodata     &       1.921   \\
                          & $\pm 0.07$  &       \nodata     &  $\pm 0.008$  \\

$m_2\sin ^3i$ (M$_\odot$) &      1.49   &       \nodata     &       1.503   \\
                          & $\pm 0.05$  &       \nodata     &  $\pm 0.004$  \\

$a_1\sin i$ ($10^6$\,km)  &   \nodata   &       \nodata     &      8.950   \\
                          &             &                   & $\pm 0.008$  \\

$a_2\sin i$ ($10^6$\,km)  &   \nodata   &       \nodata     &     11.441   \\
                          &             &                   &  $\pm 0.024$  \\

rms residual (km~s$^{-1}$) &  \nodata   &       \nodata     &      0.11   \\
(unit weight)   \\
\enddata
\tablenotetext{a}{The orbital element {\it T} usually
denotes a time of periastron in an eccentric orbit.  However,  
Kondo's {\it T}, for which he uses the symbol $T_0$,
identifies a time of the descending 
node of the primary; KKK's \citep{ketal01} value is a time of 
primary minimum since they analyzed the star as an eclipsing binary;
our value is a time of periastron passage.}
\end{deluxetable}

\clearpage

\begin{deluxetable}{lr}
\tablewidth{250pt}
\tablecaption{PHOTOMETRIC ELEMENTS AND  
PHYSICAL CHARACTERISTICS OF RR~LYN\label{tbl-6}}
\tablehead{
\colhead{Parameter}   & \colhead{Value}       \\
}
\startdata
$i$ (deg     )                    &    $87.45 \pm 0.11$  \\
$r_1$                             &    $0.0878 \pm 0.0005$ \\
$r_2$                             &    $0.0541 \pm 0.0011$ \\
$a$ ($ = a_1 + a_2$) (R$_\odot$)  &    $29.32 \pm 0.04$  \\
$m_1$ (M$_\odot$)                 &    $1.927 \pm 0.008$  \\
$m_2$ (M$_\odot$)                 &    $1.507 \pm 0.004$  \\
$R_1$ (R$_\odot$)                 &    $2.57 \pm 0.02$  \\
$R_2$ (R$_\odot$)                 &    $1.59 \pm 0.03$  \\
\enddata
\tablecomments{Inclination and fractional radii ($r_1$ and $r_2$) are 
from \citet{ketal01}.}
\end{deluxetable}

\clearpage

\begin{deluxetable}{lrrr}
\tablewidth{0pt}
\tablecaption{ORBITAL ELEMENTS OF 12~BOO\label{tbl-7}}
\tablehead{
\colhead{}    &
\colhead{}    &
\colhead{McD+KP} &
\colhead{McD+KP+CfA}      \\
\colhead{Parameter}  & \colhead{BTH\tablenotemark{a}}& \colhead{Solution}  &
\colhead{Solution}      \\
}
\startdata
$P$ (days)                                &      9.6045492  &       9.604596  &       9.6045529  \\
                                       & $\pm 0.0000076$ &  $\pm 0.000021$ &  $\pm 0.0000048$ \\

$T$ (HJD)\tablenotemark{b} & 2,451,238.2729     &  2,452,400.4266    
&   2,452,400.4292     \\
                                       & $\pm 0.0051$    &  $\pm 0.0043$   &  $\pm 0.0035$    \\

$e$                                    &      0.19233    &       0.19243   &       0.19268    \\
                                       & $\pm 0.00086$   &  $\pm 0.00047$  &  $\pm 0.00042$   \\

$\omega$ ($^\circ$)                    &    286.67       &     286.88      &     286.87       \\
                                       & $\pm 0.19$      &  $\pm 0.16$     &  $\pm 0.14$      \\

$K_1$ (km~s$^{-1}$)                    &     67.302      &      67.283     &      67.286      \\
                                       & $\pm 0.087$     &  $\pm 0.042$    &  $\pm 0.037$     \\

$K_2$ (km~s$^{-1}$)                    &     69.36       &      69.29      &      69.30       \\
                                       & $\pm 0.10$      &  $\pm 0.06$     &  $\pm 0.05$      \\

$\gamma$ (km~s$^{-1}$)                 &      9.551      &       9.588     &       9.578      \\
                                       & $\pm 0.051$     &  $\pm 0.024$    &  $\pm 0.022$     \\

$m_1\sin ^3i$ (M$_\odot$)              &     \nodata     &       1.218     &       1.218      \\
                                       &                 &  $\pm 0.002$    &  $\pm 0.002$     \\

$m_2\sin ^3i$ (M$_\odot$)              &     \nodata     &       1.183     &       1.183      \\
                                       &                 &  $\pm 0.002$    &  $\pm 0.001$     \\

$a_1\sin i$ ($10^6$\,km)               &     \nodata     &       8.720     &       8.720      \\
                                       &                 &  $\pm 0.006$    &  $\pm 0.005$     \\

$a_2\sin i$ ($10^6$\,km)               &     \nodata     &       8.981     &       8.981      \\
                                       &             & $\pm 0.008$    &  $\pm 0.007$     \\

rms residual (km~s$^{-1}$)\tablenotemark{c} &  0.49 &  0.18   &   0.18       \\

rms residual (km~s$^{-1}$)\tablenotemark{d} & \nodata & 0.10  &   0.11       \\
\enddata
\tablenotetext{a}{BTH \citep{betal05} elements are from their "full-fit" 
solution in their Table 4.}
\tablenotetext{b}{The BTH value of $T$ is incremented by 0.5 to convert
from the MJD they used to the HJD used here.}
\tablenotetext{c}{Rms residuals for solutions of McD+KP and McD+KP+CfA 
velocities are for McD and KP primary and secondary velocities.}
\tablenotetext{d}{Rms residuals for McD primary velocities, {\em i.e.} the 
velocities of unit weight.}
\end{deluxetable}

\clearpage

\begin{deluxetable}{lrr}
\tablewidth{0pt}
\tablecaption{ORBITAL ELEMENTS OF HR~6169\label{tbl-8}}
\tablehead{
\colhead{Parameter}           &
\colhead{Young 1920}        &
\colhead{This study}       \\
}
\startdata
$P$ (days)                   &     10.56    &        10.559435  \\
                          & $\pm 0.005$  &    $\pm 0.000055$ \\

$T$ (HJD)                 &  2,422,422.236   &     2,453,171.529  \\
                          & $\pm 0.046$  &    $\pm 0.004$ \\

$e$                       &      0.430   &         0.4140  \\
                          & $\pm 0.012$  &    $\pm 0.0012$ \\

$\omega$ (deg)            &      4.12    &        10.69  \\
                          & $\pm 2.66$   &    $\pm 0.20$ \\

$K_1$ (km~s$^{-1}$)       &     62.41    &        71.35  \\
                          & $\pm 1.08$   &    $\pm 0.38$ \\

$K_2$ (km~s$^{-1}$)       &    101.36    &        95.46  \\
                          & $\pm 1.64$   &    $\pm 0.13$ \\

$\gamma$ (km~s$^{-1}$)    &    $-9.88$   &      $-18.33$ \\
                          & $\pm 0.85$   &    $\pm 0.09$ \\

$m_1\sin ^3i$ (M$_\odot$) &      2.19    &         2.197 \\
                          &              &    $\pm 0.012$ \\

$m_2\sin ^3i$ (M$_\odot$) &      1.35    &         1.642 \\
                          &              &    $\pm 0.015$ \\

$a_1\sin i$ ($10^6$\,km)  &      8.18    &         9.431  \\
                          &              &    $\pm 0.051$ \\

$a_2\sin i$ ($10^6$\,km)  &     13.28    &        12.618  \\
                          &              &    $\pm 0.019$ \\

rms residual (km~s$^{-1}$) &      4.3     &         0.26  \\
(unit weight)   \\
\enddata
\end{deluxetable}





\begin{thebibliography}{}
\bibitem[Abt et al.(2002)]{aetal02}
Abt, H. A., Levato, H., \& Grosso, M. 2002, \apj, 573, 359

\bibitem[Abt \& Levy(1976)]{al76} 
Abt, H. A., \& Levy, S. G. 1976, \apjs, 30, 273

\bibitem[Abt \& Morrell(1995)]{am95}
Abt, H. A., \& Morrell, N. I. 1995, \apjs, 99, 135

\bibitem[Balachandran(1990)]{b90}
Balachandran, S. 1990, \apj, 354, 310

\bibitem[Barden(1985)]{b85}
Barden, S. C. 1985, \apj, 295, 162

\bibitem[Batten et al.(1989)]{betal89} 
Batten, A. H., Fletcher, J. M., \& MacCarthy, D. G. 1989, Publ. Dominion
Astrophy. Obs., 17, 1

\bibitem[Belserene(1947)]{b47} 
Belserene, E. P. 1947, \apj, 105, 229

\bibitem[Bikmaev et al.(2002)]{betal02}
Bikmaev, I. F., et al. 2002, \aap, 389, 537

\bibitem[Boden et al.(1999)]{betal99} 
Boden, A. F., et al. 1999, \apj, 527, 360

\bibitem[Boden et al.(2000)]{betal00}
Boden, A. F., Creech-Eakman, M., \& Queloz, D. 2000, \apj, 536, 880

\bibitem[Boden et al.(2005)]{betal05} 
Boden, A. F., Torres, G., \& Hummel, C. A.  2005, \apj, 627, 464 

\bibitem[Botsula(1960)]{b60} 
Botsula, R. A. 1960, Byull. Astron. Obs. im. \'Engel'garda, No. 35, 43

\bibitem[Budding(1974)]{b74} 
Budding, E. 1974, \apss, 30, 433

\bibitem[Colavita et al.(1999)]{cetal99} 
Colavita, M. M., et al. 1999, \apj, 510, 505

\bibitem[Cowley et al.(1969)]{cetal69}
Cowley, A., Cowley, C., Jaschek, M., and Jaschek, C. 1969, \aj, 74, 375

\bibitem[De Medeiros \& Udry(1999)]{du99} 
De Medeiros, J. R., \& Udry, S.  1999, \aap, 346, 532

\bibitem[Duquennoy \& Mayor(1991)]{dm91} 
Duquennoy, A., \& Mayor, M. 1991, \aap, 248, 485

\bibitem[Fekel(1997)]{f97}
Fekel, F. C. 1997, \pasp, 109, 514

\bibitem[Fekel \& Tomkin(2004)]{ft04} 
Fekel, F. C., \& Tomkin, J. 2004, Astronomische Nachrichten, 325, 649

\bibitem[Fitzpatrick(1993)]{f93}
Fitzpatrick, M. J. 1993, in ASP Conf. Ser. 52, Astronomical
Data Analysis Software and Systems II, ed. R. J. Hanisch, R. V. J.
Brissenden, \& J. Barnes (San Francisco: ASP), 472

\bibitem[Flower(1996)]{f96}
Flower, P. J. 1996, \apj, 469, 355

\bibitem[Gray(1992)]{g92}
Gray, D. F. 1992, The Observation and Analysis of Stellar Photospheres
(Cambridge: Cambridge University Press)

\bibitem[Gray et al.(2001)]{getal01}
Gray, R. O., Napier, M. G., \& Winkler, L. I. 2001, \aj, 121, 2148

\bibitem[Harper(1914)]{h14} 
Harper, W. E. 1914, Publ. Dominion Obs., 1, 303

\bibitem[Harper(1915)]{h15} 
Harper, W. E. 1915, Publ. Dominion Obs., 2, 167

\bibitem[Huenemoerder \& Barden(1984)]{hb84}
Huenemoerder, D. P., \& Barden, S. C. 1984, BAAS, 16, 510

\bibitem[Huffer(1931)]{h31} 
Huffer, C. M. 1931, Publ. Washburn Obs., 15, 199

\bibitem[Hummel et al.(2003)]{hetal03} 
Hummel, C. A. et al. 2003, \aj, 125, 2630

\bibitem[Hut(1981)]{h81}
Hut, P. 1981, \aap, 99, 126

\bibitem[Johnson(1966)]{j66}
Johnson, H. L. 1966, \araa, 4, 196

\bibitem[Johnson \& Morgan(1953)]{jm53}
Johnson, H. L., \& Morgan, W. W. 1953, \apj, 117, 313

\bibitem[Khaliullin \& Khaliullina(2002)]{kk02} 
Khaliullin, Kh. F., \& Khaliullina, A. I. 2002, Astron. Zh., 79, 137 
[Astron. Rep., 46, 119 (2002)] 

\bibitem[Khaliullin et al.(2001)]{ketal01} 
Khaliullin, Kh. F., Khaliullina, A. I., \& Krylov, A. V. 2001, Astron. Zh., 
78, 1014 [Astron. Rep., 45, 888 (2001)] 

\bibitem[Khaliullin et al.(1985)]{ketal85}
Khaliullin, K. Mironov, A. V., \& Moshkalyov, V. G. 1985, \apss, 111, 291

\bibitem[Konacki(2005)]{k05} 
Konacki, M. 2005, ApJ, 626, 431

\bibitem[Konacki \& Lane(2004)]{kl04} 
Konacki, M., \& Lane, B. F. 2004, \apj, 610, 443

\bibitem[Kondo(1976)]{k76} 
Kondo, M. 1976, Annals of the Tokyo Astron. Obs, 16, 1

\bibitem[Lavrov et al.(1988)]{lavetal88} 
Lavrov, M. I., Lavrova, N. V., \& Shabalov, Yu. F. 1988, Tr. Kazan. Obs., 
Issue 51, p. 19

\bibitem[Lebre et al.(1999)]{letal99}
L\`ebre, A., De Laverny, P., De Medeiros, J., Charbonnel, C., \& Da Silva,
L. 1999, \aap, 345, 936

\bibitem[Linnell(1966)]{l66} 
Linnell, A. P. 1966, AJ, 71, 458

\bibitem[Lyubimkov \& Rachkovskaya(1995)]{lr95} 
Lyubimkov, L. S., \& Rachkovskaya, 1995, Astron. Zh., 72, 64 
[Astron. Rep., 39, 56 (1995)]

\bibitem[Magalashvili \& Kumsishvili(1959)]{mk59} 
Magalashvili, N. L., \& Kumsishvili, Ya. I. 1959, Byull. Akad. Nauk Gruz. 
SSR, Abastumanskaya Astrofiz. Obs., No. 24, 13

\bibitem[Matthews \& Mathieu(1992)]{mm92}
Matthews, L. D., \& Mathieu, R. D. 1992, in ASP Conf. Ser. 32, Complimentary
Approaches to Double and Multiple Star Research, IAU Colloquium 135,
ed. H. A. McAlister \& W. I. Hartkopf (San Francisco: ASP), 244

\bibitem[McCarthy et al.(1993)]{metal93} 
McCarthy, J. A., Sandiford, B. A., Boyd, D., \& Booth, J. 1993, \pasp, 105, 881

\bibitem[Moore et al.(1966)]{metal66} 
Moore, C. E., Minnaert, M. G. J., \& Houtgast, J. 1966,
   The Solar Spectrum 2935\,\AA\ to 8770\,\AA,
   NBS monograph 61 (U.S. Government Printing Office, Washington D.C.)

\bibitem[Perryman et al.(1997)]{petal97} 
Perryman, M. A. C., et al. 1997, \aap, 323, L49

\bibitem[Petrie(1950)]{p50}
Petrie, R. M. 1950, Publ. Dominion Astrophys. Obs., 8, 319

\bibitem[Popper(1971)]{p71} 
Popper, D. M. 1971, ApJ, 169, 549

\bibitem[Pourbaix et al.(2004)]{petal04} 
Pourbaix, D. et al. 2004, \aap, 424, 727

\bibitem[Quirrenbach(2001)]{q01} 
Quirrenbach, A. 2001, Ann. Rev. Astron. Ap., 39, 353

\bibitem[Roman(1949)]{r49}
Roman, N. G. 1949, \apj, 110, 205

\bibitem[Roman(1950)]{r50}
Roman, N. G. 1950, \apj, 112, 554 

\bibitem[Roman(1952)]{r52}
Roman, N. G. 1952, \apj, 116, 122

\bibitem[Scarfe et al.(1990)]{setal90}
Scarfe, C. D., Batten, A. H., \& Fletcher, J. M. 1990, Publ. Dominion
Astrophy. Obs., 18, 21

\bibitem[Stockton \& Fekel(1992)]{sf92}
Stockton, R. A., \& Fekel, F. C. 1992, \mnras, 256, 575 

\bibitem[Strassmeier \& Fekel(1990)]{sf90}
Strassmeier, K. G., \& Fekel, F. C. 1990, \aap, 230, 389

\bibitem[Tassoul \& Tassoul(1992)]{tt92}
Tassoul, J.-L., \& Tassoul, M. 1992, \apj, 395, 259

\bibitem[Taylor(2003)]{t03}
Taylor, B. J. 2003, \aap, 398, 731

\bibitem[ten Brummelaar et al.(2003)]{tbetal03} 
ten Brummelaar, T. A. et al. 2003, Proc. SPIE,
   4838, 69

\bibitem[Traub et al.(2003)]{tetal03} 
Traub, W. A. et al. 2003, Proc. SPIE, 4838, 45

\bibitem[Young(1920)]{y20} 
Young, R. K. 1920, Publ. Dom. Astrophys. Obs., 1, 233

\bibitem[Zahn(1977)]{z77}
Zahn, J.-P. 1977, \aap, 57, 383

\end{thebibliography}
\end{document}